\documentclass[12pt]{iopart}
\usepackage{graphicx}
\usepackage{epstopdf}
\usepackage{cite}
\usepackage{iopams}
\usepackage{subfigure}

\begin{document}

\title{Evolution of Geometric Structures in Intense Turbulence}

\author{Haitao Xu$^{1,2}$\footnote{Author to whom correspondence should be addressed}, Nicholas T Ouellette$^{1,2}$\footnote{Present address: Department of Physics, Haverford College, Haverford, PA 19041, USA}, and Eberhard Bodenschatz$^{1,2,3,4,5}$}
\address{$^1$International Collaboration for Turbulence Research}
\address{$^2$Max Planck Institute for Dynamics and Self-Organization, 37077 G\"{o}ttingen, Germany}
\address{$^3$Laboratory of Atomic and Solid State Physics, Cornell University, Ithaca, NY 14853, USA}
\address{$^4$Sibley School of Mechanical and Aerospace Engineering, Cornell University, Ithaca, NY 14853, USA}
\address{$^5$Inst.~for Nonlinear Dynamics, U.~G\"{o}ttingen, 37073 G\"{o}ttingen, Germany}
\ead{haitao.xu@ds.mpg.de}

\begin{abstract}

We report measurements of the evolution of lines, planes, and volumes in an
intensely turbulent laboratory flow using high-speed particle tracking. We find
that the classical characteristic time scale of an eddy at the initial scale of
the object considered is the natural time scale for the subsequent evolution.
The initial separation may only be neglected if this time scale is much smaller
than the largest turbulence time scale, implying extremely high turbulence
levels. 

\end{abstract}

\pacs{47.27.Jv,47.27.Gs,47.27.T-}

\submitto{\NJP}
\maketitle

The transport of material by a carrier fluid is ubiquitous in both the natural
world and in engineering applications \cite{shraiman:2000}. When the carrier
flow is turbulent, the dispersion of the transported substance can be very
rapid. Turbulent flows are also extremely efficient at mixing
\cite{tennekes:1972}, since their nonequilibrium nature drives the production
of small scales and sharp gradients where diffusion can occur rapidly. To study
both transport and mixing, it is natural to work in the Lagrangian framework
where the fundamental objects are the trajectories of individual fluid elements
\cite{yeung:2002}. 

The trajectory of a single Lagrangian particle is, in general, not sufficient
for characterizing turbulent transport. Instead, knowledge of the collective
motion of groups of particles is required \cite{shraiman:2000,falkovich:2001}.
The simplest such multiparticle quantity is the growth of the relative distance
between a pair of particles \cite{sawford:2001}. We have previously measured
this relative dispersion \cite{bourgoin:2006,ouellette:2006c}, finding
excellent agreement with Batchelor's theoretical predictions
\cite{batchelor:1950}. Geometrically, the two particles in relative dispersion
define a line, and therefore a particle pair only gives one-dimensional
information about turbulent dispersion. An even deeper understanding of
turbulent transport requires the study of higher-dimensional structures: groups
of three particles, which define a plane, and of four particles, which define a
volume.  While such structures have been considered before in models and
numerical simulations \cite{chertkov:1999,pumir:2000,biferale:2005b} and in
low-Reynolds-number experiments \cite{castiglione:2001,luethi:2007}, they have
not been investigated at the high Reynolds numbers that are common in nature.

Here, we present measurements of the shape dynamics of collections of
Lagrangian particles in an intensely turbulent laboratory water flow.  We
consider lines (two particles), planes (three particles), and volumes (four
particles).  In all of these cases, the initial size of the object plays a
strong role: the time scale determined by this initial size, which we denote by
$t_0$, characterizes the experimentally observed evolution~\cite{xu:2007b}.  We
find that two particles separate superdiffusively, as is well-known in
turbulence, but that $t_0$ divides two types of separation behavior. Triangles
formed from three particles and volumes spanned by four particles also grow in
time, but assume stationary shapes after $t_0$, with triangles evolving to a
preferred set of internal angles and volumes flattening into nearly planar
structures.

Our measurements are made using optical particle tracking \cite{ouellette:2006}
in a swirling water flow between counter-rotating baffled disks, as described
in detail elsewhere \cite{ouellette:2006c}. We characterize the strength of the
turbulence with the Taylor-microscale Reynolds number $R_\lambda = \sqrt{15 u'
L} / \nu$, where $u'$ is the root-mean-square velocity, $L$ is the correlation
length of the velocity field, and $\nu$ is the kinematic viscosity; here, we
report measurements for $R_\lambda$ as high as 815. Our polystyrene tracer
particles are smaller than or comparable to the smallest scale of the
turbulence, the Kolmogorov length scale $\eta = (\nu^3 / \epsilon)^{1/4}$,
where $\epsilon$ is the mean rate of energy dissipation per unit mass, for all
Reynolds numbers studied, and faithfully follow the flow~\cite{voth:2002}.  By
modifying our tracking algorithms, we have been able to increase the lengths of
measured trajectories significantly, allowing for the study of longer-time
statistics \cite{xu:2007}.

\begin{figure}
\begin{center}
\includegraphics[width=0.45\linewidth]{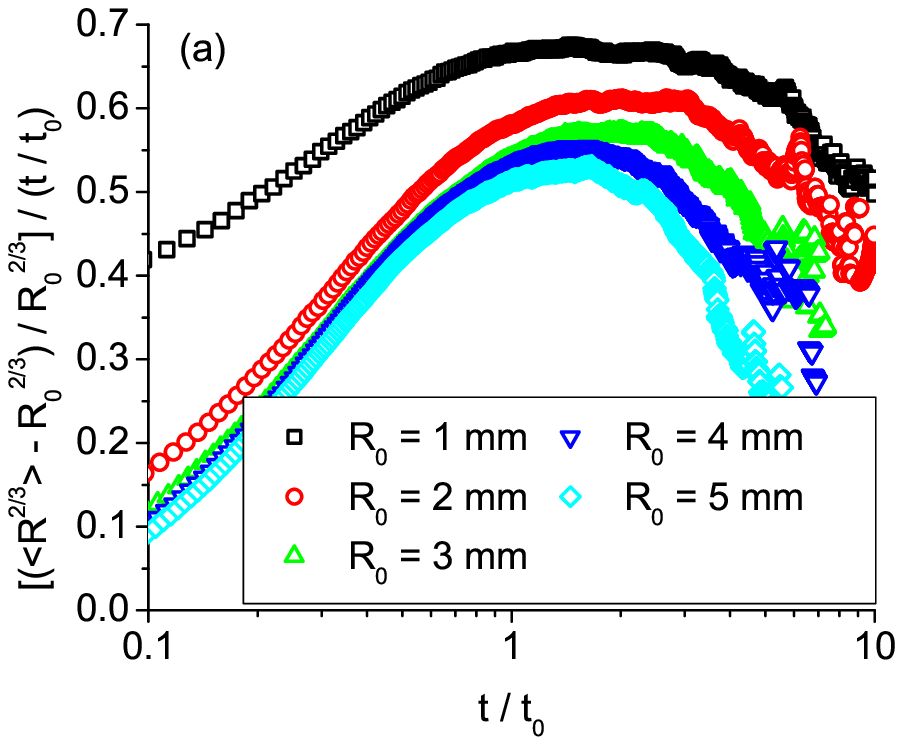}
\includegraphics[width=0.45\linewidth]{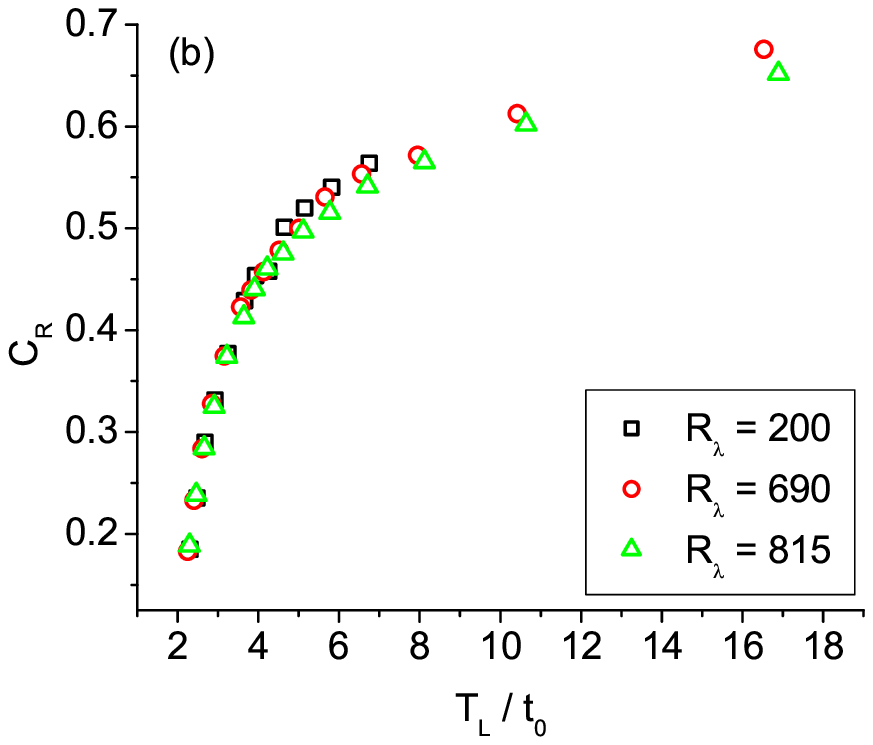}
\caption{(a) The separation of two particles in time, compared with a modified Richardson-Obukhov law, for different initial separations, ranging from 1 mm to 5 mm. The Reynolds number is fixed at $R_\lambda = 690$, with a Kolmogorov scale of $\eta = 30$ $\mu$m. We observe similar behavior at other Reynolds numbers.
(b) The change of $C_R$ with $T_L / t_0$, shown for three Reynolds numbers. }
\label{fig:line}
\end{center}
\end{figure}

Let us first consider one-dimensional shape changes by measuring the growth of the separation $\mathbf{R}(t)$ between two particles. For this case, the well-known Richardson-Obukhov law~\cite{richardson:1926,sawford:2001} predicts that
\begin{equation}
\label{eq:richardson}
\langle R^2(t) \rangle = g \epsilon t^3
\end{equation}
in the inertial range, \textit{i.e.}, $\eta \ll R \ll L$ and $\tau_\eta \ll t \ll T_L$, where $T_L = (L^2/\epsilon)^{1/3}$ is the large-eddy turnover time.
In Eq.~\eref{eq:richardson} the dimensionless coefficient $g$, known as the Richardson constant, is expected to be universal and independent of initial separation $R_0$.
It has been notoriously difficult, however, to observe conclusive evidence of this $t^3$ scaling experimentally \cite{sawford:2001}. In our previous measurements of relative dispersion \cite{bourgoin:2006,ouellette:2006c}, we instead found that the initial separation of the pair $R_0$ plays an important role, as first suggested by Batchelor \cite{batchelor:1950}. He predicted that 
\begin{equation}
\label{eq:batchelor} 
\langle \delta R_i \delta R_i \rangle = (11/3) C_2 (\epsilon R_0)^{2/3} t^2,
\end{equation}
where $\delta \mathbf{R}(t) \equiv \mathbf{R}(t) - \mathbf{R}_0$ and $C_2 = 2.13\pm 0.22$ is the scaling constant for the second-order Eulerian velocity structure function \cite{sreeni:1995}. Batchelor additionally predicted that this scaling law should hold for $t \ll t_0$, where
\begin{equation}
t_0  \equiv (R_0^2 / \epsilon)^{1/3},
\end{equation} 
may be regarded as the lifetime of an eddy of scale $R_0$. 

It has been suggested that the failure to observe the Richardson-Obukhov law is
due to the influence of particle pairs that separate anomalously slowly or
quickly, so that they bring in non-inertial-range
effects~\cite{boffetta:2002,biferale:2005a}.  This will occur unless the
inertial range is sufficiently wide and the effects of both the dissipation and
integral scales are negligible, requiring very large Reynolds numbers.  In
addition, the finite measurement volume in experiments may introduce a bias
against quickly separating particle pairs~\cite{luethi:2007a}.  We have
therefore also measured $\langle R^{2/3}(t) \rangle - R_0^{2/3}$, which is less
affected by the finite-volume bias and may display scaling behavior at Reynolds
numbers accessible in current experiments~\cite{bourgoin:2006}. If the
Richardson-Obukhov law holds, then 
\begin{equation}
(\langle R^{2/3}(t) \rangle - R_0^{2/3})/R_0^{2/3} = C_R (t/t_0), \quad (t_0 \ll t \ll T_L,  R_0 \ll R \ll L)
\end{equation}
where $C_R$ should be a constant related to the Richardson constant $g$.  The
compensated plot $(\langle R^{2/3}(t) \rangle - R_0^{2/3})/R_0^{2/3} / (t/t_0)$
should collapse to a plateau in the inertial range, independent  of initial
separation $R_0$.  As shown in Fig.~\ref{fig:line}(a), plateaus, though short,
do exist for $t \gg t_0$.  We find, however, that the initial separation $R_0$
again plays a role: $t_0$ is the time scale of the transition to the $\langle
R^{2/3} \rangle \sim t$ scaling and the plateau values depend on $R_0$.  These
observations support our earlier argument that a very large separation between
$T_L$ and $t_0$ (corresponding to a very large Reynolds number) is required to
observe $R_0$-independent Richardson-Obukhov scaling~\cite{bourgoin:2006},
which is also supported by recent work using a stochastic
model~\cite{luethi:2007a}.  To quantify the effect, we plot in
Fig.~\ref{fig:line}(b) the change of $C_R$ with $T_L/t_0$ at three different
Reynolds numbers, where $C_R$ is measured from the peak value in the plateau
region of the compensated curves. The largest $T_L/t_0$ in this plot are taken
from $R_0 \approx 30 \eta$, the lower bound of the inertial
range~\cite{pope:2000}. It can be seen that the measurements from three
different Reynolds numbers collapse very well, and that the $R_0$-independent
region is not reached even at $R_\lambda = 815$.

\begin{figure}
\begin{center}
\subfigure[]{
\includegraphics[width=0.45\linewidth]{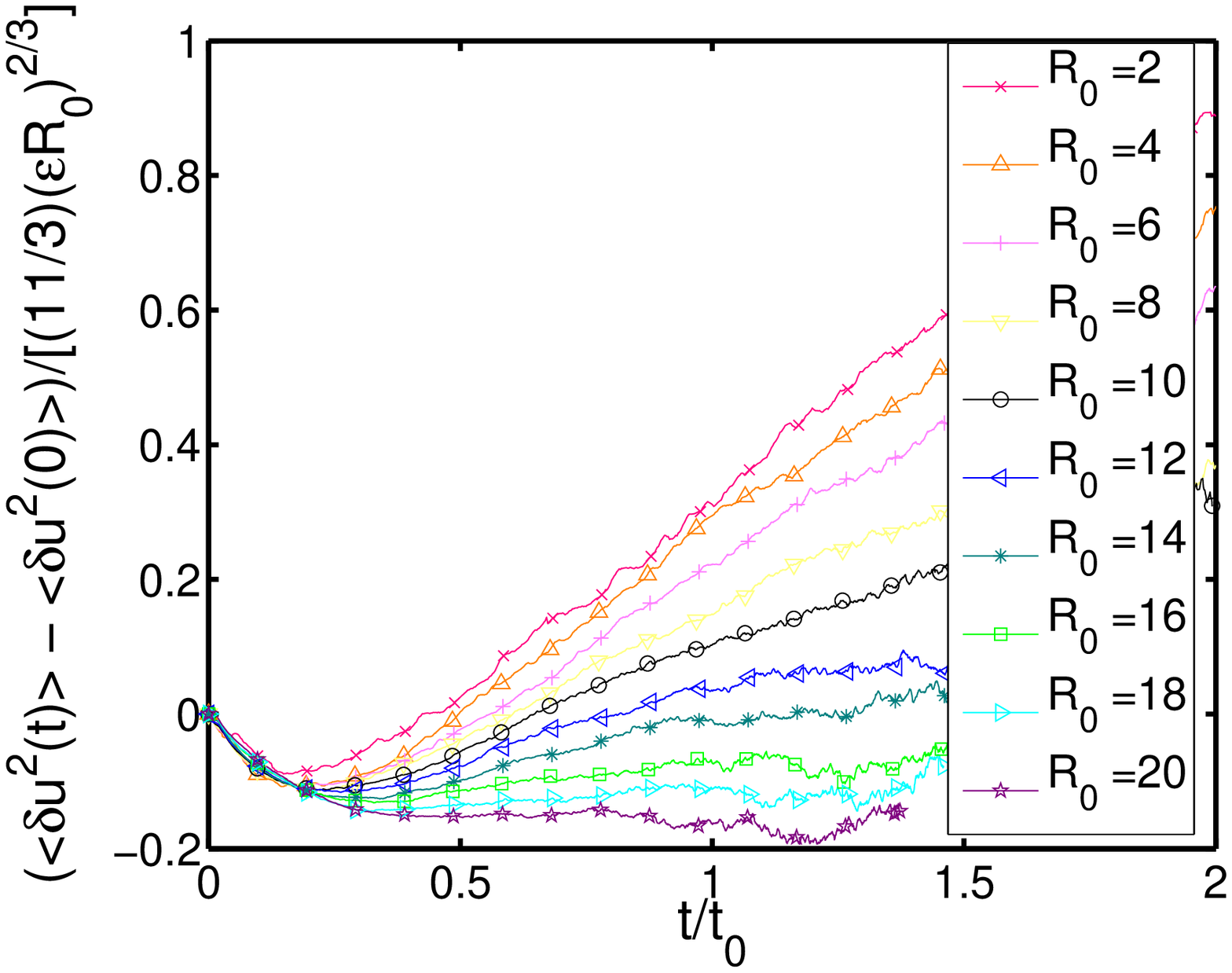}
}
\subfigure[]{
\includegraphics[width=0.45\linewidth]{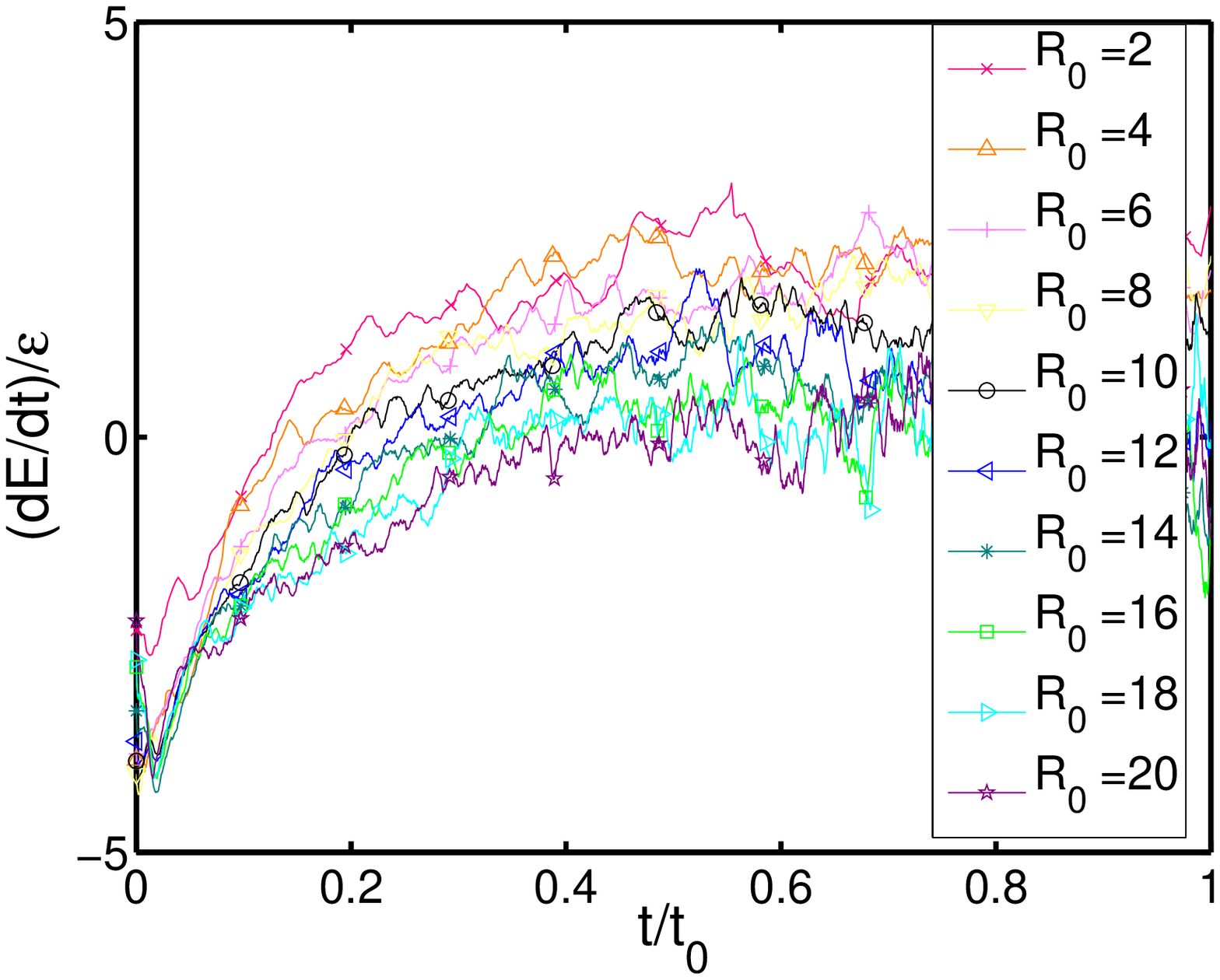}
}
\caption{(a) Change of the energy of relative motion ${\delta}u^2(t) - {\delta}u^2(0)$ following the pairs, normalized by $(11/3)(\epsilon R_0)^{2/3}$, the theoretical average relative energy at the initial separation between the pairs, where $\delta u$ is the relative velocity between the pairs. (b) The rate of energy change $d E / d t$ normalized by the turbulence dissipation rate, where $E = (\delta u^2)/2$. Both (a) and (b) are from the $R_\lambda = 690$ experiment, with a Kolmogorov scale of $\eta = 30$ $\mu$m. The legends give the initial separations in mm. We observed similar behavior at other Reynolds numbers.}
\label{fig:energy}
\end{center}
\end{figure}

The absence of Richardson universal scaling law even at $R_\lambda \sim 10^3$ may be better understood from the evolution of $\delta u$, the relative velocity between fluid particles pairs. Figure~\ref{fig:energy}(a) shows the change of the energy of relative motion, $\delta u^2(t) - \delta u^2(0)$, following the pairs. Figure~\ref{fig:energy}(b) shows the rate of the change, $d E / d t$, following the pairs, where $E \equiv \delta u^2 /2$. Clearly, the energy of relative motion decreases initially before the final increase. As initial separation increases, the return to $dE /dt >0$ takes longer time. These observations are in good agreement with results Numerical simulations~\cite{pumir:2001}, where the same change of the energy of relative motion was found for a ``cloud'' of tracer particles. This initial decrease of relative motion energy can be linked to the large correlation length of forcing in three-dimensional turbulence~\cite{falkovich:2001}. In light of Figure~\ref{fig:energy}, the relative motion between a pair of particles is first slowed down and it takes a time comparable to $t_0$ for the relative motion to accelerate again. The further apart the particles the longer the relative period of slow down is. Only after the initial decreasing period, the Richardson regime can possibly be observed.

While studying the two-particle case tells us how particles separate along
lines in turbulence, groups of three particles can show richer dynamics since
they form two-dimensional objects. We therefore measured the shape statistics
of triangles formed from three particles.  Here we report only the statistics
of triangles that were initially nearly equilateral. We then measured the
growth of the sides the triangles, analogous to the evolution of the
mean-squared separation of two particles considered above, and the distortion
of the triangles, information that is unavailable from pair separation
measurements.

\begin{figure}
\begin{center}
\includegraphics[width=0.45\linewidth]{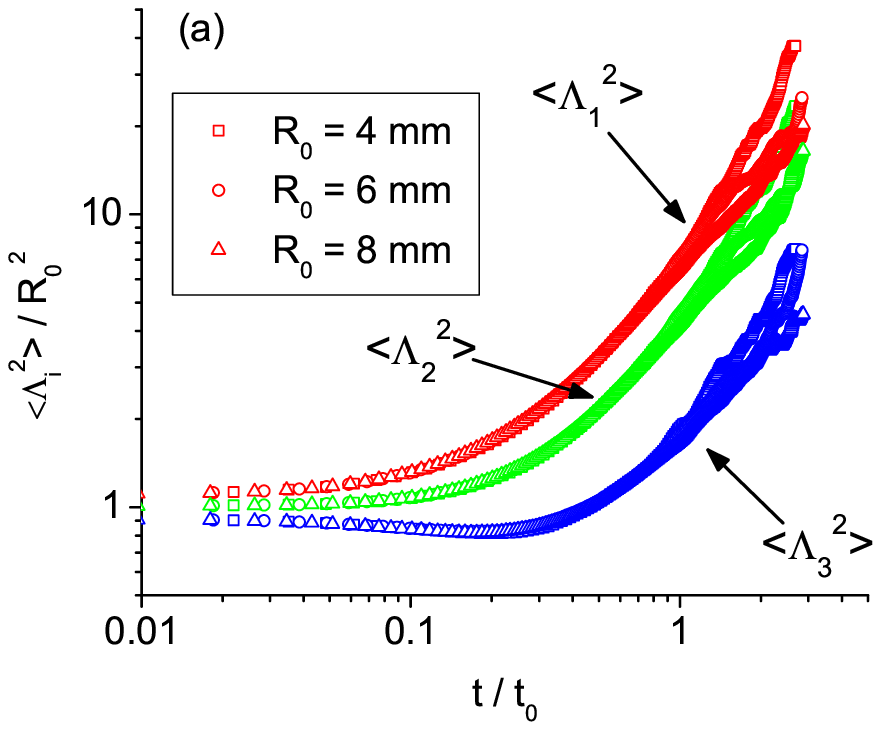}
\includegraphics[width=0.45\linewidth]{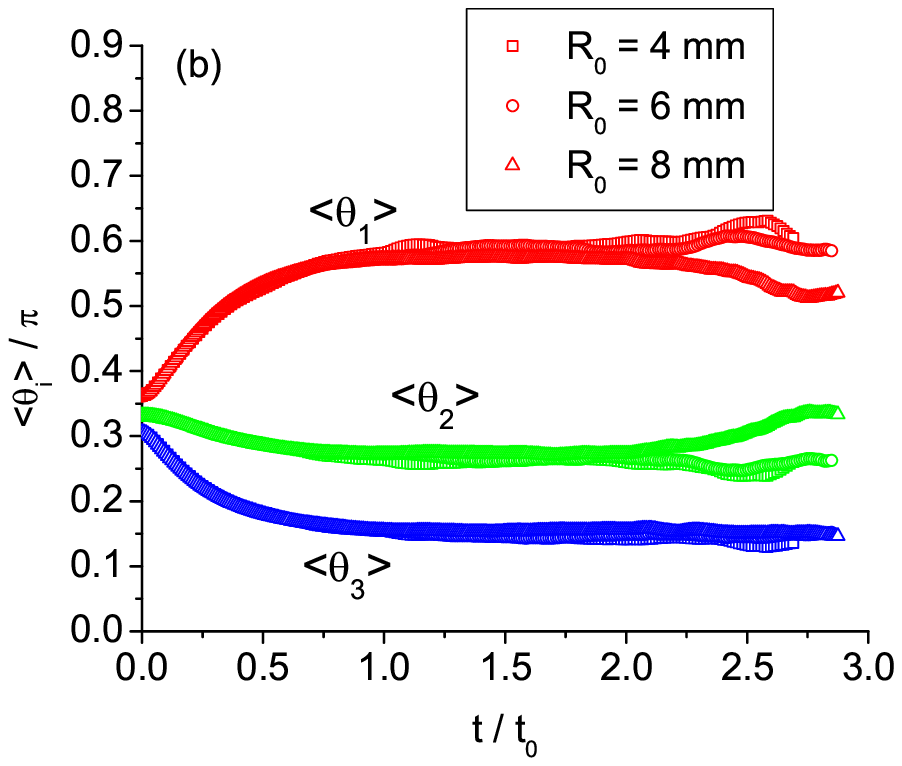}
\caption{The evolution of Lagrangian triangles at $R_\lambda = 815$ ($\eta = 23$ $\mu$m), characterized by (a) the mean-squared lengths of the triangle edges and (b) the mean triangle angles. $R_0$, the initial edge length, varies from 4 mm to 6 mm. As in the two-particle case, the data for different initial sizes collapse when time is scaled by $t_0$.}
\label{fig:tri}
\end{center}
\end{figure}

Figure~\ref{fig:tri}(a) shows the evolution of the mean-squared lengths of the
triangles $\langle \Lambda_i^2 \rangle$ (with $\Lambda_1 \geq \Lambda_2 \geq
\Lambda_3$) for three different initial triangle sizes.  When time is scaled by
$t_0$, defined now with the side length of the initially equilateral triangles,
the data collapse for different initial sizes. Additionally, scaling in this
fashion collapses data taken at different Reynolds numbers (not shown). We note
that unlike the monotonic increase of $\langle \Lambda_1^2 \rangle$ and
$\langle \Lambda_2^2 \rangle$, there is initially a slight decrease in $\langle
\Lambda_3^2 \rangle$, an effect that is absent in the statistics of particle
pairs. This result shows the importance of studying higher-dimensional
structures in three-dimensional turbulence. We shall discuss this initial
decrease in more detail when considering four-particle statistics below.

In addition to a mean growth, three particles can assume nontrivial 2D
configurations. To quantify these shapes, we show the evolution of the means of
the three triangle angles $\langle \theta_i \rangle$ in Fig.~\ref{fig:tri}(b).
Just as for the edge lengths, the angle data collapse when time is scaled by
$t_0$. We also observe that our initially equilateral triangles become
distorted over a time of order $t_0$ to stationary, obtuse shapes, with angles
of $0.56\pi$, $0.27\pi$, and $0.17\pi$. These values appear to be independent
of Reynolds number: indeed, our observations are in agreement with earlier
results from 3D numerical simulations \cite{pumir:2000} and 2D experiments
\cite{castiglione:2001}, even though they were performed at significantly lower
Reynolds number.

\begin{figure}
\begin{center}
\includegraphics[width=0.45\linewidth]{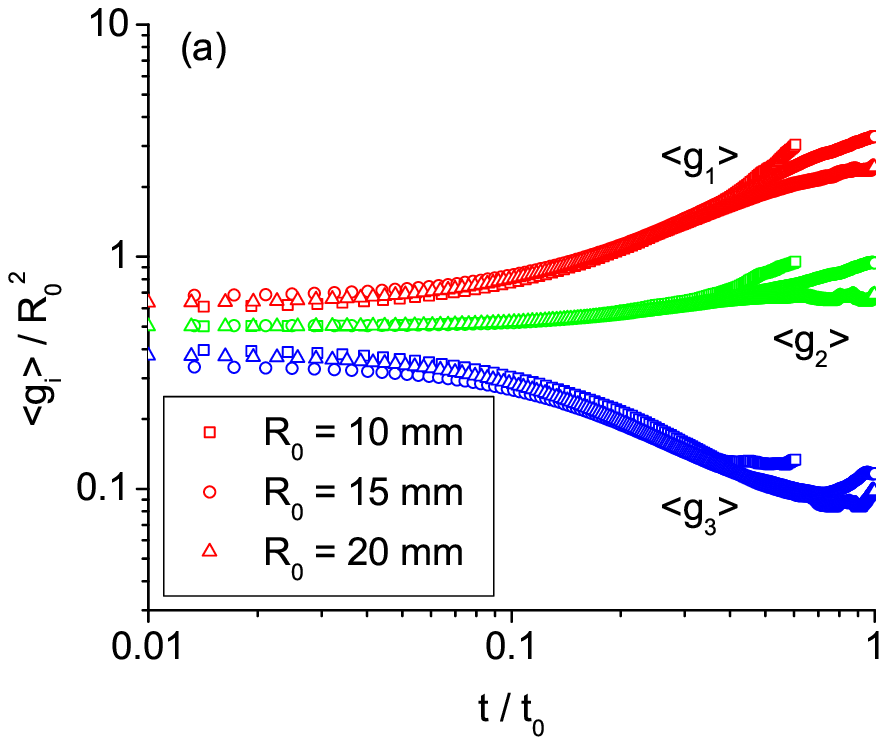}
\includegraphics[width=0.45\linewidth]{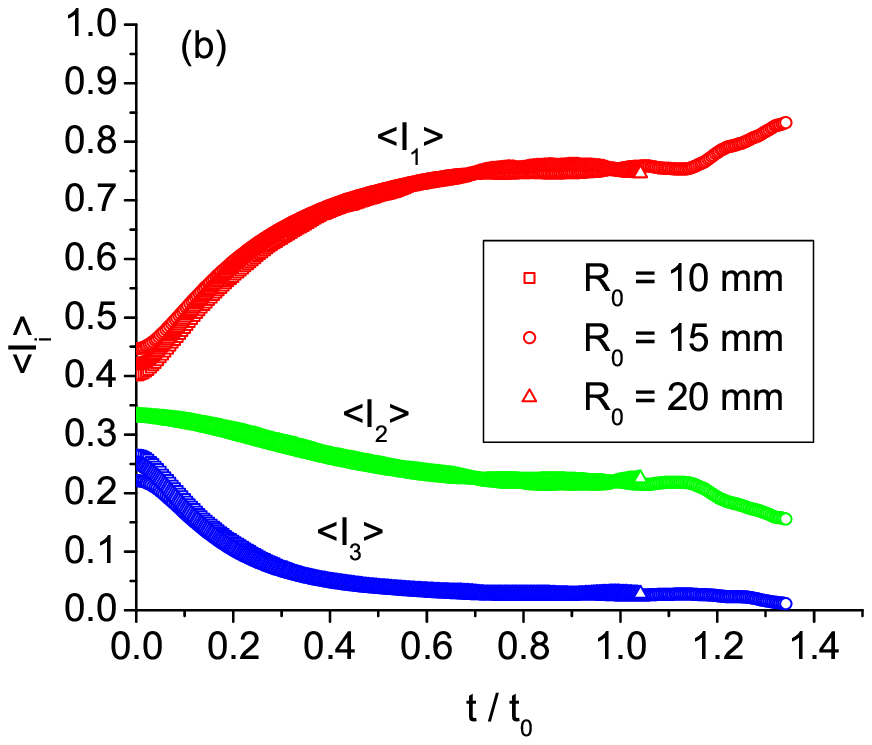}
\caption{The evolution of Lagrangian tetrads at $R_\lambda = 690$ ($\eta = 30$ $\mu$m), characterized by (a) the mean eigenvalues of the inertia tensor scaled by the initial tetrad size and (b) the mean normalized eigenvalues for initial sizes ranging from 10 mm to 20 mm. Once again, the data for different initial sizes collapse when time is scaled by $t_0$.}
\label{fig:tetrad}
\end{center}
\end{figure}

The evolution of triangles has given us more insight into transport than the two-particle case, but for 3D turbulence, we must consider volumes for a full characterization.
We therefore consider groups of four particles, the minimum number of points required to define a volume. Such groups form ``tetrads,'' which have previously been used to construct a stochastic model of the coarse-grained velocity gradient tensor \cite{chertkov:1999}. Following Chertkov \textit{et al.}~\cite{chertkov:1999},
we characterize the spatial arrangement of the particles with the eigenvalues $g_i$ (with $g_1 \geq g_2 \geq g_3$) of the ``inertia'' tensor $g_{ij} \equiv X_{ik} X_{kj}$. The column vectors of the tensor $X_{ij}$ are defined based on the relative separation between the four points:
\begin{eqnarray}
\mathbf{X}_1 &= \left( \mathbf{x}_2 - \mathbf{x}_1 \right)  / \sqrt{2}, \nonumber \\
\mathbf{X}_2 &= \left( 2 \mathbf{x}_3 - \mathbf{x}_2 - \mathbf{x}_1 \right) / \sqrt{6}, \\
\mathbf{X}_3 &= \left( 3 \mathbf{x}_4 - \mathbf{x}_3 - \mathbf{x}_2 - \mathbf{x}_1 \right) / \sqrt{12}, \nonumber
\end{eqnarray}
where $\mathbf{x}_n$ ($n = 1,2,3,4$) is the position of  the $n^{th}$ Lagrangian particle. 
The volume of the tetrad is given by $V = \left(\sqrt{g_1 g_2 g_3} \right) / 3$, and its radius of gyration is $R_g^2 = g_1 + g_2 + g_3$ \cite{pumir:2000,biferale:2005b}. We also define normalized eigenvalues $I_i = g_i / R_g^2$ to describe the tetrad shape: $I_1 = I_2 = I_3 = 1/3$ defines an isotropic tetrad, $I_3 = 0$ means that the four points are coplanar, and $I_2 = I_3 = 0$ means that they are collinear. As with the triangles studied above, we considered only tetrads that were initially nearly isotropic, with the length of all edges within 10\% of a nominal size $R_0$.

In Fig.~\ref{fig:tetrad}(a), we show the evolution of the tetrad eigenvalues
scaled by the initial tetrad size.  Data for different initial sizes collapse
when time is scaled by $t_0$, as it would for different Reynolds numbers.
Here, it is clear that while the two larger eigenvalues increase monotonically,
the smallest eigenvalue decreases. This observation provides direct
experimental evidence for earlier theoretical and numerical
work~\cite{chertkov:1999} that in 3D turbulence the large-scale, coarse grained
strain-rate tensor has, on average, two positive eigenvalues and one negative
eigenvalue. The negative eigenvalue causes the initial compression in one
direction, and consequently the decrease in $\langle g_3 \rangle$. The
normalized eigenvalues, shown in Fig.~\ref{fig:tetrad}(b), also collapse for
different sizes when scaled by $t_0$.  We also find that the tetrads develop
towards a nearly planar configuration after a time of order $t_0$. {Within the experimental resolution, we determine the}
stationary values of the normalized eigenvalues to be $\langle I_2 \rangle =
0.25$ and $\langle I_3 \rangle = 0.06$ (note that $\langle I_1 \rangle = 1-
\langle I_2 \rangle - \langle I_3 \rangle$), similar to our earlier results
where we did not select only the initially nearly isotropic tetrads
\cite{ouellette:2006d}. We note that we did not observe a Richardson-like $g_i
\sim t^3$ for any Reynolds number or initial size accessible.

\begin{figure}
\begin{center}
\subfigure[]{
\includegraphics[width=0.45\linewidth]{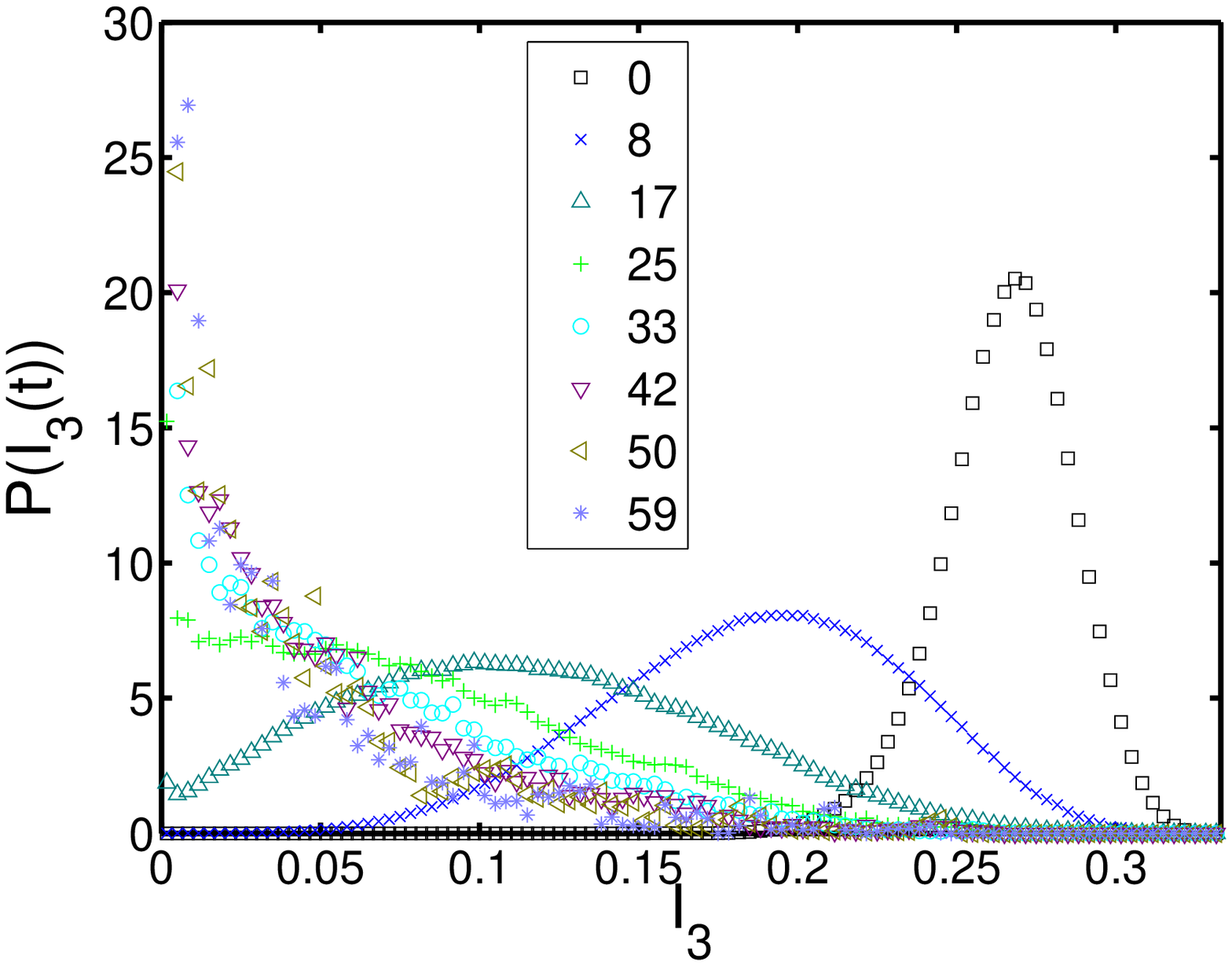}
}
\subfigure[]{
\includegraphics[width=0.45\linewidth]{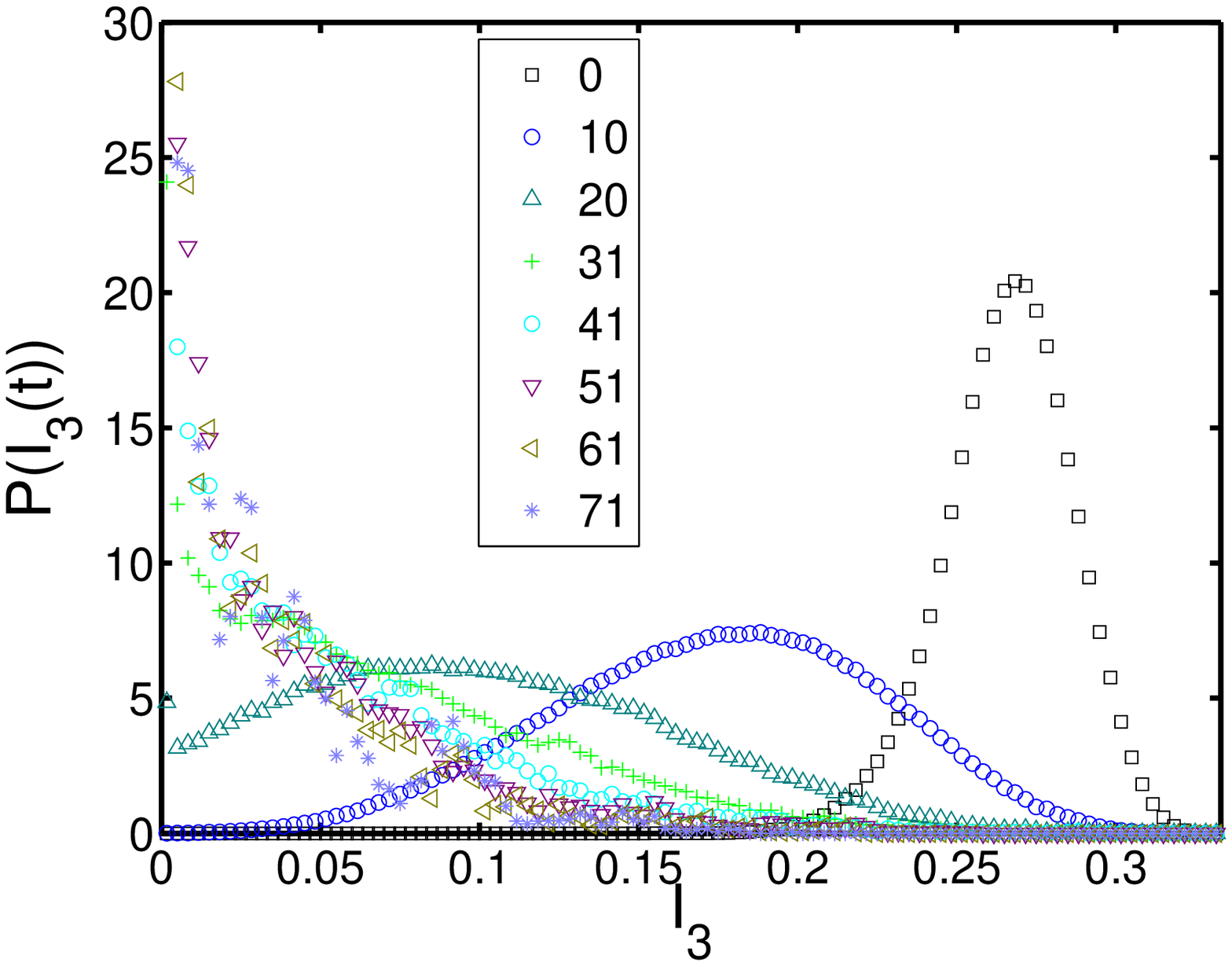}
}
\caption{The evolution of PDF of the shape factor $\langle I_3 \rangle$ following Lagrangian tetrads with initial size $R_0 = 20$mm in turbulence. The legends show the time (in units of $\tau_\eta$) at which the PDFs are measured. Similar changes are observed for tetrads with initial size of 10 and 15mm. (a) $R_\lambda = 690$ ($\eta = 30$ $\mu$m); (b) $R_\lambda = 815$ ($\eta = 23$ $\mu$m).}
\label{fig:shapePDF}
\end{center}
\end{figure}

Our results are qualitatively similar to the recent numerical simulations of
Lagrangian tetrads in Ref.~\cite{biferale:2005b}, performed at $R_\lambda =
280$, where stationary values of $\langle I_2 \rangle = 0.16$ and $\langle I_3
\rangle = 0.02$ were reported, indicating an even stronger tendency towards
planar tetrads than in our data. 
In Fig.~\ref{fig:shapePDF}, we show the evolution of the probability density function (PDF) of the smallest shape factor $\langle I_3 \rangle$ following Lagrangian tetrads. Again, the evolution of the PDFs are consistent with simulation results in Ref.~\cite{biferale:2005b} for tetrads with initial separation in the dissipative range. It can be seen that after $60\sim 70 \tau_\eta$ (or approximately $t_0$), the PDFs are still slowly evolving. Beyond that time, the tetrads, on average, have been swept away from their initial positions by half of the size of the measurement volume and disappear from our observation. Therefore, our measurement of $\langle I_i \rangle$ might be affected by the finite measurement volume. In addition, it is more probable for a vertex of an extremely deformed tetrad to move out of the finite measurement volume. There is hence a potential bias towards smaller deformation in our experiments.

A possible reason for the slight difference
between the simulation and experimental results could be the higher
potentiality for a vertex of an extremely deformed tetrad to move out of the
finite measurement volume of our experiments, at which point the tetrad would
no longer be detectable. There is therefore a potential bias towards smaller
deformation in our experiments. 

\begin{figure}
\begin{center}
\subfigure[]{
\includegraphics[width=0.45\linewidth]{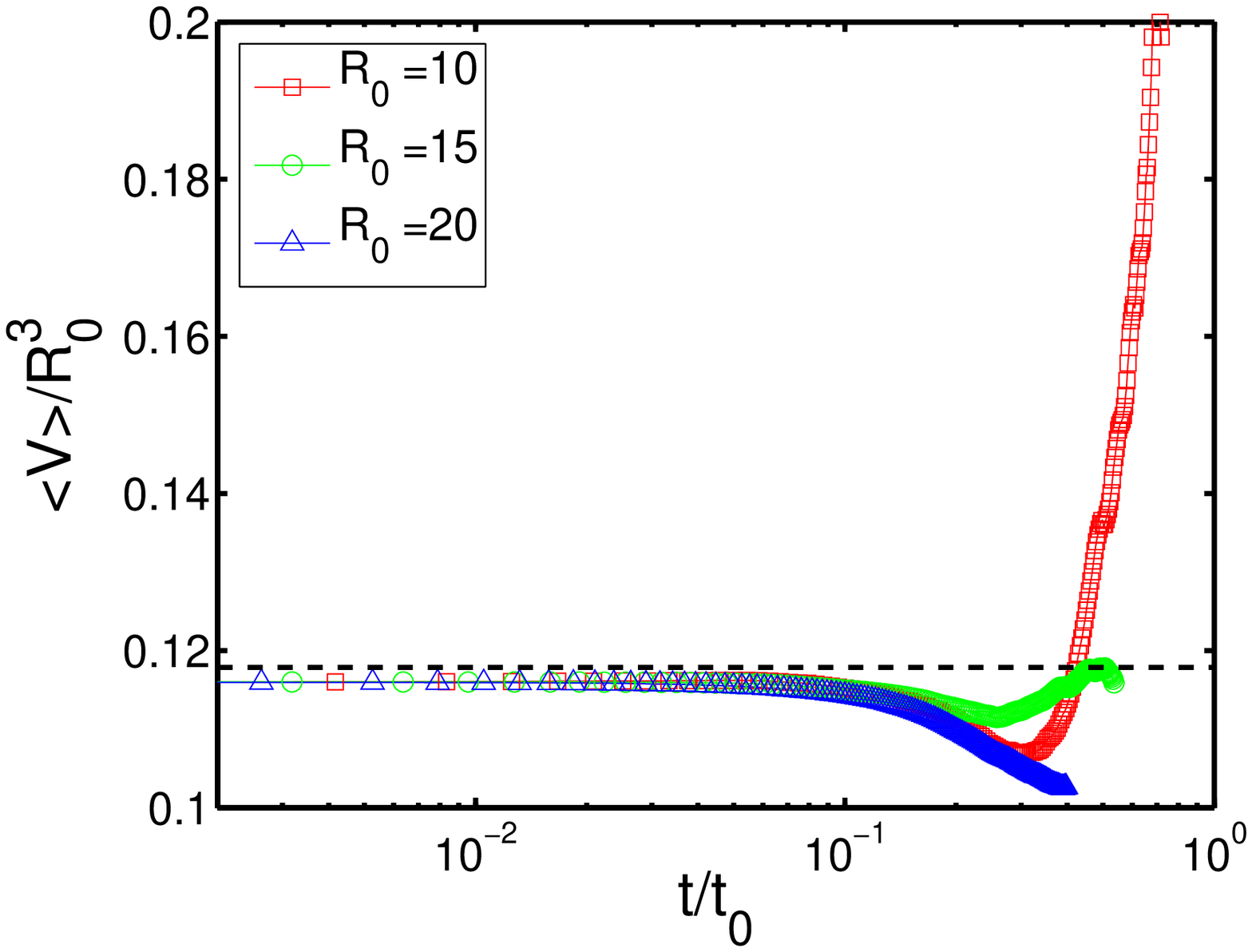}
}
\subfigure[]{
\includegraphics[width=0.45\linewidth]{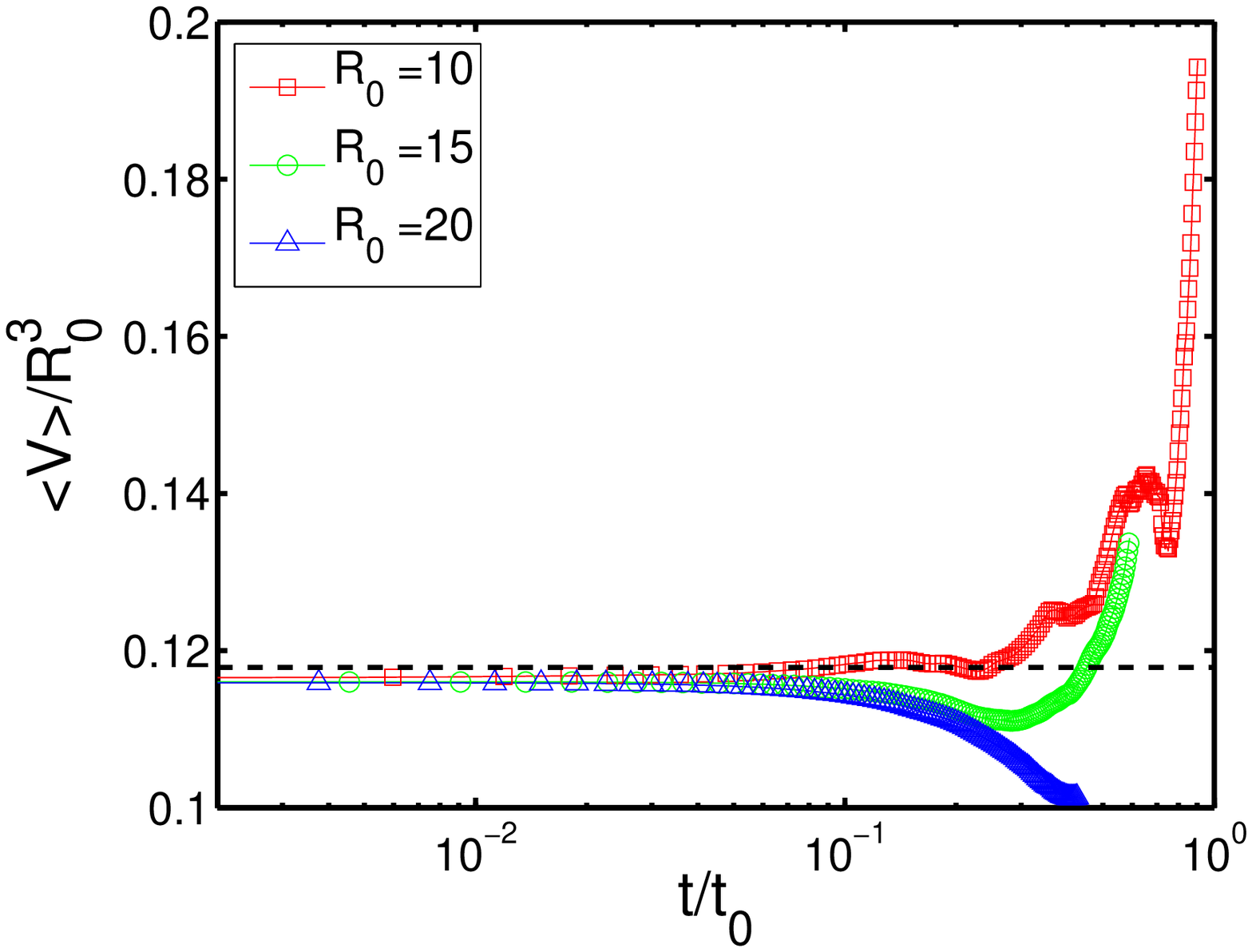}
}
\caption{The evolution of the averaged tetrad volume $\langle V \rangle$, normalized by $R_0^3$, following Lagrangian tetrads with inertial range sizes in turbulence. (a) $R_\lambda = 690$ ($\eta = 30$ $\mu$m); (b) $R_\lambda = 815$ ($\eta = 23$ $\mu$m).}
\label{fig:volume}
\end{center}
\end{figure}

The initial size of the tetrads studied in Ref.~\cite{biferale:2005b} are in
the dissipation range ($R_0 \sim \eta$), whereas our smallest initial
separation was $R_0 \sim 300 \eta$, well into the inertial range.  In
Ref.~\cite{biferale:2005b}, the initial decrease in $g_3$, similar to that
shown in Fig.~\ref{fig:tetrad}(a), was attributed to dissipative effects, which
were also presumed to be responsible for the lack of observed Richardson
scaling.  Since we observed very similar behavior with $R_0$ well in the inertial range,
however, we attribute the lack of $t^3$ scaling to a small separation between
$t_0$ and $T_L$, just as for the case of pair separation. 
The correspondence
between our inertial-range results and the smaller-scale numerical results
suggests additionally that the statistics of the coarse-grained velocity
gradient tensor \cite{chertkov:1999} may be very similar to those of the true
velocity gradient. Further study of these similarities will be an excellent
opportunity to bring together complementary experiments and numerical
simulations.

Figure~\ref{fig:volume} shows the change of averaged volume following the tetrads with different sizes. The tetrads volume remains constant initially for time $t < 0.1t_0$. Then decreases in volume before the final increase. This decrease was not observed in previous numerical work~\cite{pumir:2000} and experimental measurements~\cite{luethi:2007}. A possible reason is that the tetrad sizes are large enough to feel the effect of the mean strain.

In summary, we have measured multiparticle Lagrangian statistics in an
intensely turbulent laboratory flow. We studied the evolution of lines, planes,
and volumes, each parameterized by the minimum number of Lagrangian points.
We find that, in three-dimensional turbulence, material volumes tend to flatten into planar
shapes, in agreement with previous numerical and experimental studies at much lower Reynolds numbers. Our results clearly indicate that the initial particle separation is an
important parameter in each of these cases and must be included in models. Only
when the ratio of the largest turbulence time scale and the time scale based on
the initial size of the object is very large, implying a very large Reynolds
number, may the initial size be safely neglected.  

\ack
We thank L.~Collins, G.~Falkovich, Z.~Warhaft, and in particular A.~Pumir for many helpful discussions and suggestions.
This work was supported by the Max Planck Society and the National Science Foundation under grants PHY-9988755 and PHY-0216406.


\section*{References}
\begin{thebibliography}{10}

\bibitem{shraiman:2000}
B.~I. Shraiman and E.~D. Siggia.
\newblock Scalar turbulence.
\newblock {\em Nature}, 405:639--646, 2000.

\bibitem{tennekes:1972}
T.~Tennekes and J.~L. Lumley.
\newblock {\em A First Course in Turbulence}.
\newblock MIT Press, Cambridge, MA, 1972.

\bibitem{yeung:2002}
P.~K. Yeung.
\newblock Lagrangian investigations of turbulence.
\newblock {\em Annu. Rev. Fluid Mech.}, 34:115--142, 2002.

\bibitem{falkovich:2001}
G.~Falkovich, K.~Gaw{\c e}dzki, and M.~Vergassola.
\newblock Particles and fields in fluid turbulence.
\newblock {\em Rev. Mod. Phys.}, 73:913--975, 2001.

\bibitem{sawford:2001}
B.~L. Sawford.
\newblock Turbulent relative dispersion.
\newblock {\em Annu. Rev. Fluid Mech.}, 33:289--317, 2001.

\bibitem{bourgoin:2006}
M.~Bourgoin, N.~T. Ouellette, H.~Xu, J.~Berg, and E.~Bodenschatz.
\newblock The role of pair dispersion in turbulent flow.
\newblock {\em Science}, 311:835--838, 2006.

\bibitem{ouellette:2006c}
N.~T. Ouellette, H.~Xu, M.~Bourgoin, and E.~Bodenschatz.
\newblock An experimental study of turbulent relative dispersion models.
\newblock {\em New J. Phys.}, 8:109, 2006.

\bibitem{batchelor:1950}
G.~K. Batchelor.
\newblock The application of the similarity theory of turbulence to atmospheric
  diffusion.
\newblock {\em Q. J. R. Meteorol. Soc.}, 76:133--146, 1950.

\bibitem{chertkov:1999}
M.~Chertkov, A.~Pumir, and B.~I. Shraiman.
\newblock Lagrangian tetrad dynamics and the phenomenology of turbulence.
\newblock {\em Phys. Fluids}, 11:2394--2410, 1999.

\bibitem{pumir:2000}
A.~Pumir, B.~I. Shraiman, and M.~Chertkov.
\newblock Geometry of {L}agrangian dispersion in turbulence.
\newblock {\em Phys. Rev. Lett.}, 85:5324--5327, 2000.

\bibitem{biferale:2005b}
L.~Biferale, G.~Boffetta, A.~Celani, B.~J. Devenish, A.~Lanotte, and F.~Toschi.
\newblock Multiparticle dispersion in fully developed turbulence.
\newblock {\em Phys. Fluids}, 17:111701, 2005.

\bibitem{castiglione:2001}
P.~Castiglione and A.~Pumir.
\newblock Evolution of triangles in a two-dimensional turbulent flow.
\newblock {\em Phys. Rev. E}, 64:056303, 2001.

\bibitem{luethi:2007}
B.~L{\"u}thi, S.~Ott, J.~Berg, and J.~Mann.
\newblock Lagrangian multi-particle statistics.
\newblock {\em J. Turbul.}, 8:45, 2007.

\bibitem{xu:2007b}
H.~Xu, N.~T. Ouellette, and E.~Bodenschatz.
\newblock Multi-particle statistics--{L}ines, shapes, and volumes in high
  {R}eynolds number turbulence.
\newblock In W.-Z. Chien, editor, {\em Proc. $5^{th}$ Int. Conf. Nonlinear
  Mech.}, pages 1155--1161. 2007.

\bibitem{ouellette:2006}
N.~T. Ouellette, H.~Xu, and E.~Bodenschatz.
\newblock A quantitative study of three-dimensional {L}agrangian particle
  tracking algorithms.
\newblock {\em Exp. Fluids}, 40:301--313, 2006.

\bibitem{voth:2002}
G.~A. Voth, A.~La~Porta, A.~M. Crawford, J.~Alexander, and E.~Bodenschatz.
\newblock Measurement of particle accelerations in fully developed turbulence.
\newblock {\em J. Fluid Mech.}, 469:121--160, 2002.

\bibitem{xu:2007}
H.~Xu.
\newblock Tracking {L}agrangian trajectories in physical-velocity space.
\newblock {\em under review}, 2007.

\bibitem{richardson:1926}
L.~F. Richardson.
\newblock Atmospheric diffusion shown on a distance-neighbour graph.
\newblock {\em Proc. R. Soc. Lond. A}, 110:709--736, 1926.

\bibitem{sreeni:1995}
K.~R. Sreenivasan.
\newblock On the universality of the kolmogorov constant.
\newblock {\em Phys. Fluids}, 7:2778--2784, 1995.

\bibitem{boffetta:2002}
G.~Boffetta and I.~M. Sokolov.
\newblock Relative dispersion in fully developed turbulence: the {R}ichardson's
  law and intermittency corrections.
\newblock {\em Phys. Rev. Lett.}, 88:094501, 2002.

\bibitem{biferale:2005a}
L.~Biferale, G.~Boffetta, A.~Celani, B.~J. Devenish, A.~Lanotte, and F.~Toschi.
\newblock Lagrangian statistics of particle pairs in homogeneous isotropic
  turbulence.
\newblock {\em Phys. Fluids}, 17:115101, 2005.

\bibitem{luethi:2007a}
B.~L{\"u}thi, J.~Berg, S.~Ott, and J.~Mann.
\newblock Self-similar two particle separation model.
\newblock {\em Phys. Fluids}, 19:045110, 2007.

\bibitem{pope:2000}
S.~B. Pope.
\newblock {\em Turbulent Flows}.
\newblock Cambridge University Press, Cambridge, England, 2000.

\bibitem{pumir:2001}
A.~Pumir, B.~I. Shraiman, and M.~Chertkov.
\newblock The {L}agrangian view of energy transfer in turbulent flow.
\newblock {\em Europhys.~Lett.}, 56:379--385, 2001.

\bibitem{ouellette:2006d}
N.~T. Ouellette, H.~Xu, K.~Chang, and E.~Bodenschatz.
\newblock Statistical geometry in intensely turbulence flow.
\newblock In I.~Grant, editor, {\em Proc. $12^{th}$ Int. Sym. Flow Visual.},
  page 167. 2006.

\end{thebibliography}
\end{document}